\def\<{\langle}
\def\>{\rangle}

\documentstyle[aps,pra,multicol,epsf,tighten]{revtex}

\begin{document}
\title{Entangled Coherent State Qubits in an Ion Trap}
\author{W.\ J.\ Munro$^{1}$\cite{WJM}, G.\ J.\ Milburn$^{1}$ and
B.\ C.\ Sanders$^{2}$}
\address{$^{1}$ Department of Physics, The University of Queensland,
	Brisbane, Queensland 4072, Australia.}
\address{$^{2}$ Department of Physics, Macquarie University,
	Sydney, New South Wales 2109, Australia.}
\date{\today}
\maketitle

\begin{abstract}

We show how entangled qubits can be encoded as entangled coherent states
of two--dimensional centre-of-mass vibrational motion for two ions in
an ion trap. The entangled qubit state is equivalent to the canonical 
Bell state,
and we introduce a proposal for entanglement transfer from the two
vibrational modes to the electronic states of the two ions in order
for the Bell state to be detected by resonance fluorescence shelving methods.
\end{abstract}
\pacs{Pacs Numbers: 42.50.-p}

\begin{multicols}{2}

The qubit, or quantum bit, is the fundamental component of information
in quantum computation. In the case of a spin-$1/2$ system,
for some axis of orientation, one can identify the {\em up}
state with an {\em on} state, generally
written as~$|1\rangle_L \equiv|{\bf 1}\rangle$,
for~$\bf 1$ indicating {\em on} and the $L$ subscript indicating that
this is a logical state. The $|0\rangle_L \equiv |{\bf 0}\rangle$,
or {\em off} state, then corresponds to {\em down} for this orientation.
Qubits can thus be realised, in principle, in any spin-$1/2$ system,
such as the electronic state of a two-level atom, the polarisation
of a single photon, or the vibrational state of an ion which is
restricted to either zero- or one-phonon excitations.
The concept of qubits is useful for quantum information considerations,
but the qubit is also a useful construct for Bell inequality
tests\cite{Bel65,Cla78} and for considering the
maximally entangled canonical Bell states.

It is not necessary to restrict a qubit encoding to systems with a two
dimensional Hilbert space. For example a more exotic
form of qubit can be constructed from superpositions of
coherent states\cite{Coc99} and, as we show here,
by employing entangled coherent states\cite{San92}.
Despite both the nonorthogonality of coherent states and the unbounded
Hilbert space, Bell inequality violations are possible in both limits
$\alpha \rightarrow 0$\cite{San92} and $\alpha \rightarrow \infty$\cite{Man95},
for~$\alpha$ the dimensionless amplitude of the coherent state.
The $\alpha \rightarrow \infty$ limit is achieved by representing
the entangled coherent states in a sub-space corresponding to two
coupled spin-$1/2$ systems, and ideal canonical Bell states
are realised in the $\alpha \rightarrow \infty$ limit.
Entangled coherent states can include the entanglement of even
and odd coherent states\cite{Wie9}), which can also be
treated as coupled spin-$1/2$ systems. The advantage of entangled
even and odd coherent states, as we show, is that the
states are distinguishable by parity, so that heating which
changes the  vibrational quanta correspond to  bit
flip errors, which can be detected and  corrected
via the appropriate circuit\cite{Bra97}.

Here we show how the desired entangled coherent states can be
created for the two--dimensional centre-of-mass
vibrational mode state of two trapped ions.
This proposal involves the generalisation of experimental techniques for
generating even coherent states for the motional
state of one ion in one dimension\cite{Mon96,Win98}.
The advantage of distinguishing the logical states by phonon number parity
has been shown for the case of one--dimensional motion\cite{Coc99,Sch98a}.
We demonstrate that these entangled coherent states can be
represented as entangled qubit states, and, moreover,
such a state is equivalent to a canonical Bell state
up to unitary transformation with respect to one of the two vibrational
modes, that is up to a local unitary transformation.
In order to make measurements on the entangled coherent states we give a
procedure for swapping entanglement from the
vibrational to the internal electronic states of the ions which
can then be read by resonance shelving methods.

The two-mode coherent state
\begin{equation}
|\alpha,\beta\rangle \equiv |\alpha\rangle_a \otimes |\beta\rangle_b
\end{equation}
can be prepared in an entangled coherent state via the mutual
phase-shift interaction $H_I = \hbar \chi a^\dagger a b^\dagger b$;
this interaction has been studied in detail in the context of quantum
nondemolition measurements\cite{Als88} and for implementing phase gates for
photon qubits\cite{Tur95}. In the ion trap,  the two-mode coherent
state corresponds to a two--dimensional Gaussian wavepacket for
the centre-of-mass motion of the two trapped ions.
The mutual phase-shift interaction between these two vibrational
modes of freedom for the ion can be achieved by an appropriate Raman laser
excitation\cite{Ste97}.

After an interaction time~$t=\pi/\chi$, the output state is\cite{San00}
\begin{eqnarray}
\label{ecs}
|\psi\rangle
	&=& \frac{1}{\sqrt 2} \left( |\alpha\rangle_a \otimes |+\rangle_b
		+ |-\alpha\rangle_a \otimes |-\rangle_b \right)
			\nonumber	\\
	&=& \frac{1}{\sqrt 2} \left( |+\rangle_a \otimes |\beta\rangle_b
		+ |-\rangle_a \otimes |-\beta\rangle_b \right),
\end{eqnarray}
where the even and odd coherent states are defined by
\begin{eqnarray}\label{ecsnew}
| \pm \rangle_a
	& \equiv & {\cal N}_{\pm}(\alpha) \left( |\alpha\rangle_a
		\pm |-\alpha\rangle_a \right), \;
			\nonumber	\\
| \pm \rangle_b
	& \equiv & {\cal N}_{\pm}(\beta) \left( |\beta\rangle_b
		\pm |-\beta\rangle_b \right) ,
\end{eqnarray}
with~${\cal N}_\pm$ being the appropriate normalisation coefficients given by
\begin{eqnarray}
{\cal N}_\pm(\xi)&=&1/ \sqrt{2\pm2e^{-2|\xi|^2}}
\end{eqnarray}
We will generally ignore these normalisation coefficients unless
otherwise stated.

The state in eq~(\ref{ecs}) is equivalent, up to a local (single-oscillator)
unitary transformation, to a Bell state for a particular encoding.
Following Ref~\cite{Coc99}, the logical states are encoded in
terms of even and odd coherent states, {\em viz}
\begin{equation}
\label{logical}
|{\bf 0}\rangle \leftrightarrow |+\rangle, \;
|{\bf 1}\rangle \leftrightarrow |-\rangle,
\end{equation}
and the Discrete Fourier Transform (DFT) states are represented by
\begin{equation}
\label{dual}
\overline{|{\bf 0} \rangle} \leftrightarrow |{\bf 0}\rangle +
|{\bf 1}\rangle, \;
\overline{|{\bf 1} \rangle} \leftrightarrow |{\bf 0}\rangle -
|{\bf 1}\rangle.
\end{equation}
We can ignore normalisation coefficients and write the state~(\ref{ecs}) as
\begin{eqnarray}
\label{psi}
|\psi\rangle
	&=& \overline{|{\bf 0}\rangle}_a \otimes |{\bf 0}\rangle_b
	+ \overline{|{\bf 1}\rangle}_a \otimes |{\bf 1}\rangle_b
		\nonumber	\\
	&=& |{\bf 0}\rangle_a \otimes \overline{|{\bf 0}\rangle}_b
	+ |{\bf 1}\rangle_a \otimes \overline{|{\bf 1}\rangle}_b .
\end{eqnarray}
A single-qubit rotation on either oscillator~$a$ or~$b$ of the form
\begin{equation}
\label{qubit:rotation}
\overline{|{\bf 0}\rangle} \rightarrow |{\bf 0}\rangle, \;
\overline{|{\bf 1}\rangle} \rightarrow |{\bf 1}\rangle
\end{equation}
leads to~$|\psi\rangle$ in eq~(\ref{psi}) being in the maximally
entangled Bell state
\begin{equation}
\label{Bell}
|\phi^+ \rangle \equiv |{\bf 0}\rangle \otimes |{\bf 0}\rangle
+ |{\bf 1}\rangle \otimes |{\bf 1}\rangle.
\end{equation}

The Bell state~(\ref{Bell}) is entangled
with respect to phonon number parity.
That is, the two--dimensional oscillations are
either both in even coherent states or in odd coherent states.
A bit flip error would destroy this parity entanglement.
We now show how the prepared state~$|\psi\rangle$ in eq~(\ref{ecs})
can be transformed into the Bell state~(\ref{Bell}).

We must be able to implement the qubit rotation in the logical
basis of the mode, namely
\begin{equation}
\label{theta:rotation}
\left( \begin{array}{c}
	|\psi_0(\theta)\rangle \\ |\psi_1(\theta)\rangle
	\end{array} \right)
	= \left( \begin{array} {cc}
	\cos \theta & i \sin \theta \\
	i \sin \theta & \cos \theta
	\end{array} \right)
	\left( \begin{array} {c}
	|{\bf 0}\rangle \\ |{\bf 1}\rangle
	\end{array} \right) .
\end{equation}
We present one simple, but approximate, scheme to achieve this rotation.
For~$D(\beta) \equiv \exp (\beta a^\dagger - \beta^* a)$
the displacement operator, we can express the displaced coherent state as
\begin{equation}
D(\beta) |\alpha\rangle
	= e^{i {\rm Im}(\alpha \beta^*)} |\alpha + \beta\rangle~,
\end{equation}
which acquires a phase shift~Im$(\alpha \beta^*)$.
Displacements can be effected in ion traps via the Raman laser 
scheme\cite{Mon96,Win98}.
We assume that bosonic coding employs coherent states with real amplitudes,
and we assume that~$\epsilon \equiv - i \beta$ is real to obtain
\begin{equation}
D(i\epsilon) |\alpha\rangle
	\approx e^{ i\alpha\epsilon } |\alpha+i\epsilon\rangle .
\end{equation}

If we let~$\theta = \alpha \epsilon$ be fixed,
with~$\epsilon \rightarrow 0$ and~$\alpha \rightarrow \infty$,
then we obtain the rotation~(\ref{theta:rotation}) for
\begin{equation}
\label{rotation:displacement}
|\psi_0(\theta)\rangle \sim D(i\epsilon) |{\bf 0}\rangle, \;
|\psi_1(\theta)\rangle \sim D(i\epsilon) |{\bf 1}\rangle.
\end{equation}
The displacement--effected rotation is approximate but adequate for sufficiently
small~$\epsilon$. In order to quantify the effectiveness of this
approach to rotation, we consider the fidelity of the operation:
\begin{eqnarray}
F	&=& \left| \left\langle \psi_0(\theta) \left| D(i\epsilon)
	\right| {\bf 0} \right\rangle \right|^2
		\nonumber \\
	&=& \left| \left\langle \psi_1(\theta) \left| D(i\epsilon)
	\right| {\bf 1} \right\rangle \right|^2
		\nonumber \\
	&=& \exp (-\epsilon^2) .
\end{eqnarray}
Here we have explicitly taken the normalisation in Eq.\ (\ref{ecsnew}) into
account. The fidelity approaches unity exponentially with respect
to~$\epsilon^2$ and hence is a good approximation for small $\epsilon$.

The Bell state can thus be created for the state of the
two--dimensional vibrational mode. However, direct detection of the Bell state
is not possible with current technology. An entanglement transfer from the
vibrational mode to the internal electronic states of the ions would allow
detection of the entanglement due to the existence of the Bell state.
The electronic state of an ion can be `rotated' and read with current
technology.

In order to transfer entanglement from vibrational to electronic degrees
of freedom, we need to be able to effect the transfer
\begin{equation}
\label{transfer}
\left( c_0 |{\bf 0}\rangle + c_1 |{\bf 1}\rangle \right) |0\rangle_{\rm e}
	\rightarrow |{\bf 0}\rangle \left( c_0 |0\rangle_{\rm e} + c_1
|1\rangle_{\rm e} \right),
\end{equation}
for~$\{ |0\rangle_{\rm e}, |1\rangle_{\rm e} \}$ the two electronic
states of the ion. The transfer~(\ref{transfer})
is achieved via the swap operation
\begin{mathletters}
\begin{eqnarray}
\label{swap}
|{\bf 0}\rangle \otimes |0\rangle_{\rm e}
	& \rightarrow & |{\bf 0}\rangle \otimes |0\rangle_{\rm e}, \\
|{\bf 0}\rangle \otimes |1\rangle_{\rm e}
&	\rightarrow & |{\bf 1}\rangle \otimes |0\rangle_{\rm e}, \\
|{\bf 1}\rangle \otimes |0\rangle_{\rm e}
	 &\rightarrow &|{\bf 0}\rangle \otimes |1\rangle_{\rm e},\\
|{\bf 1}\rangle \otimes |1\rangle_{\rm e}
&	\rightarrow & |{\bf 1}\rangle \otimes |1\rangle_{\rm e} .
\end{eqnarray}
\end{mathletters}

The swap operation can be realised via a sequence of three controlled 
{\sc CNot} gates.
The vibrational qubit is the control and the electronic qubit is the target for
the first and third gates, and the reverse holds for the second qubit.
We now discuss how to realise these two types of {\sc CNot} gates.

In the first case, where the vibrational qubit is the control,
it is necessary for the electronic qubit to be prepared in the ground state
and to become excited  if and only if the vibrational qubit contains an
odd number of phonons.
% , that is
% \begin{mathletters}
% \begin{eqnarray}
% U_{\rm ve} |{\bf 0}\rangle \otimes |0\rangle_e
% 	& = &\;\;\;|{\bf 0}\rangle \otimes |0\rangle_e , \;
% 	 \\
% U_{\rm ve} |{\bf 0}\rangle \otimes |1\rangle_e
% 	& = & \;\;\;|{\bf 0}\rangle \otimes |1\rangle_e ,\;
% 		\\
% U_{\rm ve} |{\bf 1}\rangle \otimes |0\rangle_e
% 	& = &\;\;\;|{\bf 1}\rangle \otimes |1\rangle_e , \;
% 		 \\
% U_{\rm ve} |{\bf 1}\rangle \otimes |1\rangle_e
% 	& = & \;\;\;|{\bf 1}\rangle \otimes |0\rangle_e.
% \end{eqnarray}
% \end{mathletters}
This transformation is achieved via the unitary
operator
\begin{equation}
\label{Ucp}
U_{\rm ve} = \exp \left[-i \pi a^\dagger a \sigma_y \right] \times
\exp \left[i \pi  a^\dagger a |1\rangle_e \langle 1|\right],
\end{equation}
which can be achieved by employing several Raman pulses at the carrier
frequency. Schneider {\it et. al}\cite{Sch98b} explicitly considered
a unitary operator of the form  $\exp \left[-i \pi a^\dagger a
\sigma_z \right]$. Noting that $\sigma_{y}=U
\sigma_{z} U^{\dagger}$, where $U$ is a single qubit rotation,
the $\exp \left[-i \pi a^\dagger a \sigma_y
\right]$ operator can be achieved by first applying a single
qubit rotation to the electronic state and then by performing the
$\exp \left[-i \pi a^\dagger a \sigma_z \right]$ operation via Raman
pulses.

The $\exp \left[-i \pi a^\dagger a \sigma_y
\right]$ part of the unitary operator (\ref{Ucp}) transforms the 
input states as follows:
\begin{mathletters}
\begin{eqnarray}
\exp \left[-i \pi a^\dagger a \sigma_y \right] |{\bf 0}\rangle 
\otimes |0\rangle_e
	& = &\;\;\;|{\bf 0}\rangle \otimes |0\rangle_e , \;	 \\
\exp \left[-i \pi a^\dagger a \sigma_y \right] |{\bf 0}\rangle 
\otimes |1\rangle_e
	& = & \;\;\;|{\bf 0}\rangle \otimes |1\rangle_e ,\;		\\
\exp \left[-i \pi a^\dagger a \sigma_y \right]|{\bf 1}\rangle \otimes 
|0\rangle_e
	& = &-|{\bf 1}\rangle \otimes |1\rangle_e , \;		 \\
\exp \left[-i \pi a^\dagger a \sigma_y \right] |{\bf 1}\rangle 
\otimes |1\rangle_e
	& = & \;\;\;|{\bf 1}\rangle \otimes |0\rangle_e.
\end{eqnarray}
\end{mathletters}
Whereas the operator $\exp \left[i \pi  a^\dagger a  |1\rangle_e \langle 1|\right]$
flips the sign of the $|{\bf 1}\rangle \otimes
|1\rangle_e$ term
\begin{eqnarray}
\exp \left[i \pi  a^\dagger a  |1\rangle_e \langle 1|\right]
|{\bf 1}\rangle \otimes |1\rangle_e
	& = & \;\;\;-|{\bf 1}\rangle \otimes |1\rangle_e,
\end{eqnarray}
while leaving the other states
$|{\bf 0}\rangle \otimes |0\rangle_e$, $|{\bf 0}\rangle \otimes |1\rangle_e$ and $|{\bf 1}\rangle \otimes
|0\rangle_e$ unchanged.
Hence the unitary transformation (\ref{Ucp})
is a {\sc CNot} with the vibrational modes being the control bit and
the electronic mode the target.

The second {\sc CNot} gate reverses the roles of the vibrational and 
electronic qubits.
Therefore, phonon number parity must be changed if the ion is in the
excited state.
% , that is
% \begin{mathletters}
% \begin{eqnarray}
% U_{\rm ev} |{\bf 0}\rangle \otimes |0\rangle_e
% 	& = &  \;\;\;|{\bf 0}\rangle \otimes |0\rangle_e , \;
% 	 \\
% U_{\rm ev} |{\bf 0}\rangle \otimes |1\rangle_e
% 	& = & \;\;\;|{\bf 1}\rangle \otimes |1\rangle_e ,\;
% 	 \\
% U_{\rm ev} |{\bf 1}\rangle \otimes |0\rangle_e
% 	& = &  \;\;\;|{\bf 1}\rangle \otimes |0\rangle_e , \;
% 	\\
% U_{\rm ev} |{\bf 1}\rangle \otimes |1\rangle_e
% 	& = & \;\;\;|{\bf 0}\rangle \otimes |1\rangle_e ,
% \end{eqnarray}
% \end{mathletters}
The required unitary transformation is the conditional displacement of the
vibrational mode if and only if the ion is in the excited state,
and such conditional displacements have been achieved
experimentally\cite{Mon96,Win98}. The corresponding unitary operator is
\begin{equation}\label{D1}
U_{\rm ev}
	= \exp \left[ i\epsilon \left( a + a^\dagger \right)
	|1\rangle_e \langle 1| \right] \exp \left[ -i \pi |1\rangle_e \langle
	1|/2 \right] ,
\end{equation}
with~$\theta=\alpha \epsilon = \pi/2$.
The $\exp \left[ i\epsilon \left( a + a^\dagger \right)
	|1\rangle_e \langle 1| \right]$ part in (\ref{D1})
gives
\begin{mathletters}
\begin{eqnarray}
\exp \left[ i\epsilon \left( a + a^\dagger \right)
	|1\rangle_e \langle 1| \right] |{\bf 0}\rangle \otimes |0\rangle_e
	& = &  \;\;\;|{\bf 0}\rangle \otimes |0\rangle_e , \;	 \\
\exp \left[ i\epsilon \left( a + a^\dagger \right)
	|1\rangle_e \langle 1| \right] |{\bf 0}\rangle \otimes |1\rangle_e
	& = & -|{\bf 1}\rangle \otimes |1\rangle_e ,\;	 \\
\exp \left[ i\epsilon \left( a + a^\dagger \right)
	|1\rangle_e \langle 1| \right] |{\bf 1}\rangle \otimes |0\rangle_e
	& = &  \;\;\;|{\bf 1}\rangle \otimes |0\rangle_e , \;	\\
\exp \left[ i\epsilon \left( a + a^\dagger \right)
	|1\rangle_e \langle 1| \right] |{\bf 1}\rangle \otimes |1\rangle_e
	& = &-|{\bf 0}\rangle \otimes |1\rangle_e ,
\end{eqnarray}
\end{mathletters}
while the second term $\exp \left[ -i \pi |1\rangle_e \langle 1|/2
\right]$ flips the sign of the $ |{\bf 0}\rangle \otimes |1\rangle_e$
and $|{\bf 1}\rangle \otimes |1\rangle_e$ states. Hence the unitary
operator (\ref{D1}) performs the  required {\sc CNot} operation with
the electronic mode as the control and the vibrational mode as the
target.

It is straightforward to then show that the sequence
\begin{equation}
\label{Uswap}
U_{\rm swap} = U_{\rm ve} U_{\rm ev} U_{\rm ve}
\end{equation}
produces the desired entanglement swap. This sequence should be
achievable with current experimentally technology.

In current ion trap experiments heating of the vibrational mode,
though small, cannot be neglected. A simple model of heating for
a vibrational mode with annihilation operator $a$
is described by the master equation\cite{schneider}
\begin{eqnarray}
\frac{d\rho}{dt}& = & \frac{\gamma}{2}\left (2a^\dagger \rho a+2a\rho
a^\dagger\right .\\
& &\;\;\;\;\; \left . -(a^\dagger a+aa^\dagger)\rho
	- \rho(a^\dagger a+a^\dagger a)\right ).
\end{eqnarray}
  This master equation describes two conditional point processes; one
corresponds to an upward  transition in phonon number
at rate $\gamma\langle aa^\dagger\rangle$ and the other to a downward
transition at the rate $\gamma\langle a^\dagger a\rangle$.
For the states discussed in this paper these two rates are
approximately the same, at least initially.
The mean value of the amplitude does not decay,
but the average energy increases at the constant rate~$\gamma$.
We can thus model the heating by two independent jump processes.
We will assume that over each run of the experiment,
taking time $\tau$, the heating rate is low enough that we only need to
consider at most a single jump, either up or down,
with probability $\delta=\gamma|\alpha|^2T$.
If only a single jump occurs, no matter which
way ( upwards or downwards), it flips the parity of the state.
In other words,  heating leads to bit-flip errors.

Up to a fidelity of $\exp (-\epsilon^2)$,
the pure Bell state~$\rho =  {1\over 2} |\phi^+\rangle \langle \phi^+|$
is obtained. In order to test a Bell inequality with the the entangled
coherent states, a large number of runs of the
experiment would need to be performed,
and the state may vary from one run
to the next if bit flip errors occur. Thus the test
of the Bell inequality is actually performed on a mixed state.  Provided
that a time interval~$\tau$ is chosen such that the
probability of more than one bit-flip error due to heating is negligible,
the density matrix for the state can be expressed as
\begin{equation}
\label{rho:bitflip}
\rho = {1\over 2}(1-\delta) |\phi^+\rangle \langle\phi^+|
	+{1\over 2}\delta |\psi^+\rangle \langle\psi^+| ,
\end{equation}
with
\begin{equation}
|\psi^+\rangle \equiv |{\bf 0}\rangle \otimes |{\bf 1}\rangle
		+ |{\bf 1}\rangle \otimes |{\bf 0}\rangle
\end{equation}
being another of the four maximally entangled Bell states.
The Bell state~$|\psi^+\rangle$ is orthogonal
to the desired state $|\phi^+\rangle$.
%For~$\gamma$ the loss rate to the environment and $\alpha$
%sufficiently large, the density matrix~$\rho$
%is correct with~$\delta = \gamma |\alpha|^2 \tau$.

The  state given by (\ref{rho:bitflip}) for $\delta$ sufficiently
small must violate the spin Bell inequality\cite{Bel65,Cla78}
\begin{eqnarray}
\label{spinbell}
B	&=& \Big| E\left(\theta_{1},\theta_{2}\right)
	+E\left(\theta_{1},\theta_{2}'\right)
	\nonumber	\\	&& \;\;\;\;\;\;\;\;\;\;\;
	+E\left(\theta_{1}',\theta_{2}\right)
	-E\left(\theta_{1}',\theta_{2}'\right)\Big| \leq  2
\end{eqnarray}
where the correlation function $E\left(\theta_{1},\theta_{2}\right)$
is given by the expectation value
\begin{eqnarray}\label{correlation}
E\left(\theta_{1},\theta_{2}\right)
	&=&\left\langle \sigma^{(1)}_{\theta_{1}}
\sigma^{(2)}_{\theta_{2}}\right\rangle ~.
\end{eqnarray}
Here the operator $\sigma^{(i)}_{\theta_{i}}$ may be defined as
\begin{equation}
\sigma^{(i)}_{\theta_{i}}=\cos \theta_{i} \,
\sigma^{(i)}_x + \sin \theta_{i} \, \sigma^{(i)}_y~
\end{equation}
where the operators $\sigma_a^{(i)}$ (with $a=x$, $y$ or $z$)
are the $a$ Pauli spin operators for the two--level system of atom $i$.
The tunable parameters~$\theta_i$ ($i=1,2$) control the
proportion of  $\sigma^{(i)}_x$ to  $\sigma^{(i)}_y$ in
$\sigma^{(i)}_{\theta_{i}}$ and function like variable polarisers in
the single--photon experiments.
% n($\theta_i$ is a adjustable analyser setting).

In an ion trap this correlation function (\ref{correlation})
is achieved by first applying a single qubit rotations to both ions and
then by performing a simultaneous measurement of $\hat \sigma_{z}$
on both ions. The $\hat \sigma_{z}$ measurement is achieved with high precision
via the shelving fluorescence technique\cite{shelving}.
The experiment is repeated over many runs and the
average gives the desired correlation function
$E\left(\theta_{1},\theta_{2}\right)$. Mathematically this correlation
function can be expressed in the form
\begin{eqnarray}
E\left(\theta_{1},\theta_{2}\right)&=& {\rm Tr} \left[\rho\;\;
\hat{V}_1^{1/2}(\theta_{1}) \;\sigma_{z}^{(1)}\;\left(\hat{V}_1^{1/2}
(\theta_{1})\right)^{\dagger}\right. \nonumber \\
&&\;\;\;\;\;\;\;\;\;\;\;\;\left. \times
\;\hat{V}_2^{1/2}(\theta_{2})\; \sigma_{z}^{(2)}\;
\left(\hat{V}_2^{1/2}(\theta_{2})
\right)^{\dagger} \right]
\end{eqnarray}
where the Cirac and Zoller single--qubit rotations\cite{CZ95}
$\hat V_{i}^{k}(\phi)$ on the $i^{\rm th}$ ion are given by
\begin{eqnarray}
	\hat V_{i}^{k}(\phi) = \exp \left[-i k {\pi \over 2}\left(|1\>_{i}
	\<0| e^{-i
	\theta_{i}}+|0\>_{i} \<1| e^{i \theta_{i}} \right) \right]~.
\end{eqnarray}
This single qubit rotation is achieved by applying a carrier pulse of
length $k \pi$ with a phase $\theta_{i}$.
% In practice the measurement of the correlation function (\ref{correlation})
% is achieved by applying a single qubit rotation 
% $\hat{V}_1^\frac{1}{2}(\theta_{1})$
% to the first ion, a single qubit rotation $\hat{V}_2^\frac{1}{2}(\theta_{2})$
% to the second ion and then performing a simultaneous measurement of $\sigma_{z}$
% on both ions. The experiment is repeated over many runs and the
% average gives the desired correlation function. The $\sigma_{z}$
% measurement is achieved with high precision via the shelving fluorescence
% technique\cite{shelving}.

Returning to the density matrix given by Eq.\ (\ref{rho:bitflip}) it is
easily shown that the correlation function (\ref{correlation}) has the
simplified form
\begin{equation}\label{simply}
E\left(\theta_{1},\theta_{2}\right)= (1-\delta)\cos
\left(\theta_{1}+\theta_{2} \right)+\delta \cos
\left(\theta_{1}-\theta_{2} \right);
\end{equation}
hence, the spin Bell inequality (\ref{spinbell}) for optimal angles
choices\cite{angles} reduces to
\begin{equation}
B=2 \sqrt 2 (1-\delta) .
\end{equation}
% or
% \begin{equation}
% B=2 \sqrt 2 \delta .
% \end{equation}
A violation of this inequality is possible for $B>2$, when
$\delta<1-1/\sqrt{2}$.
Whereas the Bell inequality is technically violated,
this does not present a loophole--free
test due to the limited temporal separation of the ions.
It does however completely close the detection loophole.

To summarise, we have described how entangled qubits can
be encoded as entangled coherent states of two--dimensional
centre-of-mass vibrational motion for two ions in an
ion trap. The entangled qubit state is equivalent to the canonical Bell state,
and by transferring the entanglement from the two vibrational modes to the
electronic states of the two ions, the Bell state can be detected by resonance
fluorescence shelving methods.\\

\noindent {\em Acknowledgement:}
Part of this work was carried out at the Isaac Newton Institute for
Mathematical Sciences in Cambridge and we acknowledge their support
and hospitality as well as the support by the European Science
Foundation.
We acknowledge also the financial support of the Australian Research Council.
\\

\end{multicols}


\begin{references}
\bibitem{angles} For the correlation function given by
Eq.\ (\ref{simply}), an optimal--angle choice which maximises
the violation for small~$\delta$ is given by:
$\theta_{1}=\pi/4$, $\theta_{1}'=-\pi/4$,
$\theta_{2}=0$ and $\theta_{2}'=\pi/2$.
\end{references}
\end{document}